\documentclass[a4paper]{article}
\usepackage{graphicx}
\usepackage{amsmath}
\usepackage{fullpage}
\usepackage{graphicx}
\usepackage{float}
\usepackage{array}
\usepackage[ruled, vlined, linesnumbered, noresetcount]{algorithm2e}
\DeclareMathOperator*{\argmax}{argmax}
\SetKwRepeat{Do}{do}{while}
\usepackage[table]{xcolor}
\usepackage{multirow}

\usepackage{mathptmx}

\begin{document}
\title{Algorithmic upper bounds for graph geodetic number}
\author{
Ahmad T.~Anaqreh
\and
Bogl\'arka G.-T\'oth
\and
Tam\'as Vink\'o
\and 
\\
University of Szeged, Institute of Informatics,  Hungary 
\\
\{ahmad, boglarka, tvinko\}@inf.u-szeged.hu
}
\date{}
\maketitle
\begin{abstract}
Graph theoretical problems based on shortest paths 
are at the core of research due to their theoretical importance and applicability.
This paper deals with the geodetic number which is a global measure for simple connected graphs and it belongs 
to the path covering problems: what is the minimal-cardinality set of vertices, such that all shortest paths between its elements cover every vertex of the graph.
Inspired by the exact 0-1 integer linear programming formalism from the recent literature, we propose a new methods to obtain upper bounds for the geodetic number in an algorithmic way.
The efficiency of these algorithms are demonstrated on a collection of structurally different graphs.
\end{abstract}

\section{Introduction}
\label{sec:defs}
Path covering problems play an important role both in theory and applications mostly by reasons of their straight
interpretability. Complex notions of cover are possible, one of the most relevant set of such problems is involving shortest paths.
The graph geodetic number, which belongs to this problem set, was introduced in \cite{Harary93}.
There might be several application fields for the problem, perhaps the most straight-forward one is given in \cite{Manuel}
which poses it as a social network problem.
It turns out that calculating the geodetic number is an NP-hard problem for general graphs \cite{atici}.
As for many similar graph theoretical problems, an integer linear programming (ILP) formulation is possible and such model
was given in a recent paper by Hansen and van Omme \cite{hansen}, containing also the first computational experiments on a set of
random graphs of moderate size.
Motivated by these results, this paper empirically investigates upper bound algorithms, which, according to our experiments provide results of small gap on the same set of random graphs using relatively low computational time even on graphs with 150 nodes, as well as on real-world graphs of large scale.

In the rest of this section the definition of the geodetic number problem is given, followed by the 0-1 linear programming formalism
from \cite{hansen}. Then, in Section \ref{sec:ub} the algorithmic descriptions of the two proposed upper bound procedures are given.
Finally, we report our extensive computational experiments in Section \ref{sec:num} to demonstrate the efficiency of these algorithms.


\paragraph{Problem description.}
A simple connected graph is denoted by $G = (V,E)$, where $V$ is the set of vertices and $E$ is the set of edges.
Assume that $n = |V|$ and $m = |E|$.
Given $i,j\in V$, the set $I[i,j]$ contains all $k\in V$ which lies on any shortest paths (\emph{geodetics}) between $i$ and $j$.
The union of all $I[i,j]$ for all $i,j \in S \subseteq V$ is denoted by $I[S]$, which is called as 
\emph{geodetic closure} of $S\subseteq V$. Formally
$$
I[S] := \{ k\in V : \exists i,j \in S, k\in I[i,j]\}.
$$
The \emph{geodetic set} is a set $S$ for which $V = I[S]$. The \emph{geodetic number} of $G$ is
$$
g(G) := \min\{ |S| : S\subseteq V \text{ and } I[S] = V\}.
$$

\paragraph{A 0-1 integer linear programming model.}
In \cite{hansen} a binary integer linear programming model has been proposed which is as follows.
Let $d:V \times V \rightarrow R$ be a function to obtain the length of shortest paths.
For each node $k \in V$ define the set
$$
P_k := \{(i, j) \in V\times V \mid d(i, k) + d(k, j) = d(i, j)\}.
$$
Set $P_k$ hence contains all pairs of vertices for which a shortest path is going through node $k$. 

The 0-1 LP model is
\begin{equation}
\min \sum_{k=1}^n x_k
\label{eq:obj}
\end{equation}
subject to
\begin{alignat}{6}
1-x_k &\leq  \sum_{(i, j) \in P_k} y_{ij} &\qquad \forall\ k, i, j\in V, i < j& \label{eq:c1}\\
 y_{ij} & \leq \ x_i &\qquad \forall\ i, j\in V, i < j&\label{eq:c2}\\
 y_{ij} & \leq \ x_j &\qquad \forall\  i, j\in V, i < j& \label{eq:c3}\\
 x_i + x_j - 1 & \leq y_{ij} &\qquad \forall\  i, j\in V, i < j& \label{eq:c4}
\end{alignat}
and the variables are all binary
\begin{eqnarray*}
&x_i &\in \{0,1\} ~~~ \forall i \in V \\
&y_{ij} &\in \{0,1\} ~~~ \forall i , j \in V, i<j
\end{eqnarray*}
The value of variable $x_k$ indicates if vertex $k$ belongs to the set $S$ or not.
Thus, the sum of these binary variables needs to be minimized.
The auxiliary variables $y_{ij}$ denote the bilinear term $x_ix_j$, so \eqref{eq:c1} means $x_k$ can be 0 only if there are $i,j\in S$ such that $k$ is on their shortest path. McCormick conditions \eqref{eq:c2}$-$\eqref{eq:c4} describe the $y_{ij}=x_ix_j$ correspondence.

\section{Upper bounds}
\label{sec:ub}
The trivial upper bound for the geodetic number is $g(G)\leq n$, which is tight for the complete graph.
Chartrand \emph{et al}.~\cite{chartrand} prove that $g(G) \leq n - d + 1$, where $d$ is the length of the longest shortest
path (called \emph{diameter}) of $G$.
Other upper bounds are also given \cite{bresar,soloff,wang} but these are concerning specific graph structures.

In this section we aim at deriving upper bounds for general graphs in algorithmic ways. Two upper bound algorithms are given. The first one uses Floyd's algorithm, while the second algorithm is based on Dijkstra's algorithm.

\subsection{Greedy algorithm}
\label{sec:greedy}
The main idea of the algorithm is to generate a geodetic set in a greedy fashion: in every step choose vertex $i$ to be included in $S$ which makes the largest increase in $I[S]$. Namely, where the cardinality $\mid I[S\cup\{i\}] \setminus I[S] \mid$ is maximal.

\paragraph{Initialization.}
The initialization part is shown in Algorithm \ref{alg:GInit}.
In the lines \ref{distb}$-$\ref{diste} all-pair shortest paths are calculated, which are then used in lines \ref{Ib}$-$\ref{Ie} where 
the sets $I_{ij}$ are defined. At this point these sets are essentially the same as $I[i,j]$ which got defined already in Section \ref{sec:defs}, i.e., 
containing all nodes which lie on any shortest paths between nodes $i$ and $j$.
We use different notation here on purpose, as the sets $I_{ij}$ are subject to change later in our algorithm.
In line \ref{initS} the nodes with degree at most one are put into (the initially empty) geodetic set $S$ as these nodes must be part of it.
The geodetic closure $I[S]$ gets initialized for $S$ in lines \ref{ISb}$-$\ref{ISe}. Note that this set could be empty here.
Finally, in lines \ref{Iijb}$-$\ref{Iije} all of the already covered vertices from the sets $I_{ij}$ are removed.

\SetKwProg{Fn}{Function}{}{}
\DontPrintSemicolon
\begin{algorithm}\caption{Greedy algorithm - Initialization}
\label{alg:GInit}
\SetKwFunction{GreedyInit}{GreedyInit}
\Fn{\GreedyInit}{

$S = \emptyset, I[S] = \emptyset, I_{ij} = \emptyset \quad \forall i,j\in V$\label{empty}\;
$d_{ij}=1 \quad \forall (i,j)\in E, \ d_{ij}=\infty \quad \forall (i,j)\notin E$\\
\For(\tcp*[f]{distance calculation}){$\forall k \in V,i \in V, j \in V$\label{distb}}{ 
	$d_{ij} = \min\{d_{ij},d_{ik}+d_{jk}\}$ \tcp*[f]{by Floyd algorithm}\label{diste}}

\For(\tcp*[f]{calculation of $I_{ij}$}){$\forall i \in V, j \in V, k \in V$\label{Ib}}{
	\If{$ d_{ij}=d_{ik}+d_{kj}$}{
		$I_{ij} = I_{ij} \cup \{k\}$\label{Ie}
	}
}

$S=\{k\in V \mid \mathrm{deg}(k)\leq 1 \}$\tcp*[f]{unreachable nodes from all paths must be in S}\label{initS}

\For(\tcp*[f]{build I[S] for S}){$\forall i \in S, j \in S$\label{ISb}}{ $I[S] = I[S] \cup I_{ij}$\label{ISe}}

\For(\tcp*[f]{Update $I_{ij}$-s by removing}){$\forall i \in V, j \in V$\label{Iijb}}{ $I_{ij} = I_{ij} \setminus I[S] $\tcp*[f]{all covered vertices}\label{Iije}}
}
\end{algorithm}

\paragraph{Auxiliary functions.}
The description of our greedy method continues in Algorithm \ref{alg:LI},
where the functions {\tt LargestIncrease} and {\tt LargestIncreasePair} are defined. These compute
the vertex (and vertex pair, respectively) which increases most the covered set $I[S]$ if it is included in $S$.
Two notations are used: the sets $I_i[S]$ contain vertices which are covered if node $i$ was included in set $S$,
and sets $I_{ij}[S]$ contain vertices which are covered if both $i$ and $j$ were included in set $S$. 
The sets $I_i[S]$ are initialized as empty in line \ref{Iiempty}.
In lines \ref{IiSb}$-$\ref{IiSe} the sets $I_i[S]$ are constructed by the nodes currently in $I_{ij}$, where $i$ is not element of $S$ and
$j$ is in $S$. Then, in line \ref{ell} we define $\ell$ to be the node from $V\setminus S$ for which $I_i[S]$ is the largest.
Similarly, in the function {\tt LargestIncreasePair(V,S,I)}, at lines \ref{IijSb}$-$\ref{IijSe} the sets $I_{ij}[S]$ are calculated with nodes which would make $I[S]$ increasing if the pair $(i,j)$
were included. 
Finally, in line \ref{kh} such pair of nodes is selected for which the set $I_{ij}[S]$ is the largest.

\begin{algorithm}[t]
\label{alg:LI}
\caption{Greedy algorithm - Auxiliary functions}
\SetKwFunction{LI}{LargestIncrease}
\SetKwFunction{LIP}{LargestIncreasePair}
\Fn{\LI{$V,S,I$}}{

\For(\tcp*[f]{compute $I_i[S]$, the set increasing}){$\forall i \in V\setminus S$\label{IiSb}}{
		$ I_i[S] = \emptyset$\tcp*[f]{$I[S]$ if $i$ is included}\label{Iiempty}
		
		\For{$\forall j \in S$}{ $ I_i[S] = I_i[S] \cup I_{ij} $}\label{IiSe}}

$\ell = \displaystyle\argmax_{i \in V\setminus S} |I_i[S]|$ \label{ell} \tcp*[f]{find the node for which $I[S]$ would grow most}

\Return $\ell, I_\ell[S]$}

\medskip

\Fn{\LIP{$V,S,I$}}{ 
\DontPrintSemicolon
\For(\tcp*[f]{compute $I_{ij}[S]$, the set increasing}){$\forall i \in V\setminus S, j 
\in V\setminus S$\label{IijSb}}{
		$I_{ij}[S] = I_{ij} \cup I_i[S] \cup I_j[S]  $   \tcp*[f]{$I[S]$ if $i,j$ are included}\label{Iijempty}
\label{IijSe}}

$(k,h) = \displaystyle\argmax_{i,j \in V\setminus S} |I_{ij}[S]|$ \tcp*[f]{find pair of node for which $I[S]$ would grow most}\label{kh}

\Return $k,h,I_{kh}[S]$}
\end{algorithm}

\paragraph{Main algorithm.}
The main loop of our greedy approach is shown in Algorithm \ref{alg:3}.
The condition in line \ref{cond} checks if there is a node which is not covered yet.
In lines \ref{heurb}$-$\ref{heure} a heuristic rule is applied, which simply checks if the size of set $I_\ell[S]$ is at least 
half than the size of set $I_{kh}[S]$. This is a greedy choice,
however, other conditions could also be applied here to decide whether one or a pair of nodes should be added to $S$. 
After this, in lines \ref{Iij-up-b}$-$\ref{Iij-up-e} the sets $I_{ij}$ get updated by removing all the covered vertices from them.
In line \ref{LIcall} we need to choose again the node for which the increase of $I[S]$ is the largest.
Finally, the execution of lines \ref{AddOneCond}$-$\ref{LIPcall} depend on the parameter $AddOne$.
If it is set as $AddOne=0$, then the algorithm chooses the best pair of nodes to be added to the set $S$
by calling the {\tt LargestIncreasePair} function. Otherwise, i.e., in case of $AddOne=1$, this function call is simply skipped, 
the settings done in line \ref{consistency} are needed for keeping the consistency.
In the numerical experiments reported in Section \ref{sec:num} this simplified version will be referred as {\tt AddOne}.

Note that due to the fact that the input graph $G$ is undirected, similarly to the 0-1 LP model description in Section \ref{sec:defs},
one can assume the condition $i < j$ for all appropriate cases. This is done in the actual implementation of the greedy algorithm to 
make it faster, but we omit these details for the easier understanding in the pseudocodes.

\begin{algorithm}
\label{alg:3}
\caption{Greedy algorithm - Main}
\DontPrintSemicolon
\GreedyInit \tcp*[f]{initialize $I_{ij}$ sets and $S$}

$[\ell,I_\ell[S]] = \LI(V,S,I)$\tcp*[f]{$\ell$ would make $I[S]$ grow most}

$[k,h,I_{kh}[S]] = \LIP(V,S,I)$ \tcp*[f]{$k$ and $h$ would make $I[S]$ grow most}

\While(\tcp*[f]{the set is not geodetic yet}){$|I_\ell[S]|+|I_{kh}[S]|>0$\label{cond}}{
	\eIf(\tcp*[f]{balance adding one or two vertices to $S$}){$|I_\ell[S]|>|I_{kh}[S]|/2$\label{heurb}}{
		$S = S \cup l$\;
		$I[S] = I[S] \cup I_\ell[S]$\tcp*[f]{update $I[S]$}}
		{$S = S \cup \{k,h\}$\;
		$I[S] = I[S] \cup I_{kh}[S]$\tcp*[f]{update $I[S]$}\label{heure}}

		\For(\tcp*[f]{Update $I_{ij}$-s by removing}){$\forall i \in V, j \in V$\label{Iij-up-b}}{ $I_{ij} = I_{ij} \setminus I[S] $\tcp*[f]{all covered vertices}\label{Iij-up-e}}		

		$[\ell,I_\ell[S]] = \LI(V,S,I)$\tcp*[f]{recompute $I_\ell[S]$}\label{LIcall}
		
		\eIf(\label{AddOneCond}\tcp*[f]{$AddOne$ is a control parameter}){AddOne}{

		$k=h=0; I_{kh}[S]=\emptyset$\label{consistency}
		}{
        $[k,h,I_{kh}[S]] = \LIP(V,S,I)$ \tcp*[f]{recompute $I_{kh}[S]$\label{LIPcall}}}

		}

\end{algorithm}

\subsubsection{Computational complexity}
The greedy heuristic uses Floyd's algorithm to calculate the distances in the input graph, which needs time $\mathcal{O}(n^3)$. 
There are nested loops used to build $I_{ij}$, the complexity for these loops is again time $\mathcal{O}(n^3)$. 
The calculation of $I[S]$, as well as the update of $I_{ij}$, has time $\mathcal{O}(n^2)$.

In the main loop of the algorithm
there are nested loops, starting with an outer loop to check if there is 
still any non-empty $I_\ell[S]$ in which it can enter up to $n$ times in the worst case.
Then, the inner loop to update all $I_{ij}$ takes time $\mathcal{O}(n^2)$.
Both of the auxiliary functions to find the vertex or vertices that makes $I[S]$ growing the most, have basically two loops, resulting in time $\mathcal{O}(n^2)$.
Taking the outer and inner loop together makes the complexity of this part in total time $\mathcal{O}(n^3)$. 
Therefore the computational complexity for the heuristic algorithm is $\mathcal{O}(n^3)$.

\subsection{Locally greedy algorithm}

In our locally greedy algorithm, the purpose is the same as earlier, i.e., to find upper bound on $g(G)$ efficiently. Furthermore, by using local information only we aimed at making the algorithm faster compared to the method introduced in Section \ref{sec:greedy}. Therefore, instead of calculating all shortest paths using Floyd’s algorithm, we calculate the distances from a specific node $v$ to all nodes not in the geodetic set $S$ by Dijkstra’s algorithm.

 The details of locally greedy algorithm is shown in Algorithm \ref{alg:4}. 
 The algorithm takes node $v$ as an input, which is either a degree-one node or a simplicial node. The degree-one node is the node connected by one edge in the graph, while the simplicial node is the node that its neighbors form a clique (a complete subgraph), namely, every two neighbors are adjacent. Choosing $v$ to be degree-one node or simplicial node is based on the fact that degree-one nodes and simplicial nodes are always part of the geodetic set, see \cite{Harary93,Hossein}.
 
 Node $v$ is the starting node for set $S$. Then {\tt LargestLocalIncrease} function in line \ref{dijk} returns the node $u$ for which the set $I[S]$ would grow most. This function is detailed in lines \ref{LLI1}-\ref{LLI2}. First, it calculates the distance from node $v$ to all other nodes in the graph together with the shortest paths and fills the sets $I_{v,:}$.
Dijkstra's algorithm is not detailed in the pseudocode as it is well-known.
In lines \ref{IjS1}-\ref{IjS2} the function computes the sets $I_j[S]$ for all $j\in V\setminus S$. In line \ref{argmaxIjS} the function finds the node $u$ that would increase $I[S]$ most by adding it to $S$. 

 In lines \ref{Su}-\ref{Rupdate} the algorithm adds $u$ to the geodetic set $S$, updates $I[S]$, and then removes $I[S]$ from $R$, which is the set of remaining nodes to be covered. The main loop of the algorithm shown in lines \ref{while}-\ref{endwhile}, in each step the algorithm checks if set $R$ is empty, namely, if there are still any uncovered nodes. As long as there are still uncovered nodes the algorithm will repeat the execution in lines \ref{LLI3} by calling {\tt LargestLocalIncrease} function to node $w$.

\begin{algorithm}
\label{alg:4}
\caption{Locally Greedy algorithm }
\SetKwFunction{Dijk}{LargestLocalIncrease}
\SetKwFunction{Dijkstra}{Dijkstra}
\SetKwInOut{Input}{Input}
\SetKwProg{Fn}{Function}
\DontPrintSemicolon
\SetKwFunction{}{}
\setcounter{AlgoLine}{0}
\Input{$v$ a degree-one or simplicial node}
$R=V,S = v, I[S] = \emptyset, I_i[S] = \emptyset \quad \forall i \in V, I_{ij} = \emptyset \quad \forall i,j\in V$\\
$[u,I_u[S]]=\Dijk(v,V,S,I)$\label{dijk}\tcp*[f]{$I[S]$ would grow most for node $u$}\\
$S = S \cup \{u\}$\tcp*[f]{update $S$}\label{Su}\\
$I[S] = I[S] \cup I_u[S]\cup \{v,u\}$\tcp*[f]{update $I[S]$}\\
$R = R \setminus I[S]$\tcp*[f]{update $R$ by removing covered nodes}\label{Rupdate}\\
$w=u$\\	
\While(\tcp*[f]{the set is not geodetic yet}){$|R|>0$}
{\label{while}
	    $[u,I_u[S]] = \Dijk(w,V,S,I)$\tcp*[f]{compute $I_u[S]$}\label{LLI3}\\
	    $S = S \cup \{u\}$\tcp*[f]{update $S$}\\ 
		$I[S] = I[S] \cup I_u[S] \cup \{u\}$ \tcp*[f]{update $I[S]$}\\
		$R = R \setminus I[S]$\tcp*[f]{update R by removing covered nodes}\\
		$w=u$
}\label{endwhile}

\Fn{\Dijk{$v,V,S,I$}}
{\label{LLI1}   
$ I_{v,:} = \Dijkstra{v}$ \tcp*[f]{shortest paths by Dijkstra  algorithm}

\For()
{\label{IjS1}
$\forall j \in V\setminus S$}
       {
        \For{$\forall i \in S$}
        { 
		$ I_j[S] = I_j[S] \cup I_{ij}$
		}
}\label{IjS2}
$u = \displaystyle\argmax_{j \in V\setminus S} |I_j[S]\setminus I[S]|$\tcp*[f]{find the node for which $I[S]$ would grow most}\label{argmaxIjS}\\
\Return $u, I_u[S]$\\\label{LLI2}
}

\end{algorithm}
\subsubsection{Computational complexity}
\label{sec:comp_loc}
Locally greedy algorithm uses Dijkstra's algorithm to calculate the distances in the input graph, which needs time $\mathcal{O}(n^2)$.
There are nested loops used to build $I_j[S]$, the complexity for these loops is again $\mathcal{O}(n^2)$. In total, the \texttt{LargestLocalIncrease} function needs time $\mathcal{O}(n^2)$.
The main while loop of the algorithm is used to fill the geodetic set $S$ by calling the function \texttt{LargestLocalIncrease} until set $R$ becomes empty, that is, at most $n$ times. Thus, the total complexity is $\mathcal{O}(n^3)$, like for the other greedy methods proposed in Section \ref{sec:greedy}.

\section{Numerical experiments}
\label{sec:num}
In order to investigate the execution time and quality of the upper bound algorithms discussed in Section \ref{sec:ub},
the ILP model and the algorithms were implemented in AMPL \cite{ampl}. As a solver we used Gurobi with parameters setup {\tt mipfocus=1, timelim=3600}. Tests were run on a computer with a 3.10Ghz i5-2400 CPU and 8GB memory.

The obtained results are reported in Tables \ref{table:er}, \ref{table:ws}, \ref{table:ba} and \ref{table:real}. The meaning of the columns are: 
\begin{description}
\item{graph:} size of the graphs as number of vertices ($n$) and number of edges ($m$);
\item{exact:} exact geodetic number (or best solution found by Gurobi in case of running out of time) and the time in seconds to find this solution;
\item{greedy:} upper bound found by the greedy algorithm (i.e., Algorithm \ref{alg:3}) by using $AddOne=0$ and its execution time in seconds;
\item{greedy ({\tt AddOne}):} upper bound found by the greedy algorithm (i.e., Algorithm \ref{alg:3}) using $AddOne=1$, i.e., the addition of only 1 vertex at a time, and its execution time in seconds;
\item{locally greedy:} upper bound found by the locally greedy algorithm (i.e., Algorithm \ref{alg:4}) and its execution time in seconds.
\end{description}

\subsection{Graph instances}
As input graphs for the algorithms, 
we have generated several random graphs which are structurally different to obtain deeper insights how the solution of the geodetic number problem depends on the graph $G$. Besides, a small set of real-world graphs were also included in the benchmark.

\subsubsection{Random graphs}
The random graphs were generated by using the following standard models.
\begin{description}
\item{Erd\H os-R\'enyi (ER) model \cite{er}}, where a graph is chosen uniformly at random from the set of all graphs which have $n$ nodes and $m$ edges.

\item{Watts-Strogatz (WS) model \cite{ws}}, which produces graphs with small world properties, namely (i) the average value of all shortest paths in
the graph is low, and (ii) high clustering coefficient. Property (ii), which measures the average probability that two neighbors of a node are themselves neighbor of each other, makes them different from ER graphs.

\item{Barab\'asi-Albert (BA) model \cite{ba}}, which creates graphs using preferential attachment growing mechanism,
where the more connected a node is, the more likely it is to receive new links. This leads to scale-free property, 
i.e., power law distribution of the form $p_k \sim k^{-\alpha}$, where $p_k$ is the fraction of nodes with degree $k$
and $\alpha$ is a parameter typically in the range $2 < \alpha < 3$. Note that BA graphs have low clustering coefficient (similarly to ER graphs)
and short path lengths (similarly to WS graphs).
\end{description}
Regarding the number of nodes and edges the following approach were used:
\begin{itemize}
\item the number of nodes were $n = 10, 20, 30, 40, 50, 60, 70, 80, 100$, and
\item for the number of edges we followed the scheme as in \cite{hansen}:
\begin{itemize}
\item for each case one can have maximum $n\cdot (n-1)/2$ edges,
\item and we took 20\%, 40\%, 60\% and 80\% of this maximum number of edges.
\end{itemize}
\item Apart from these graphs, we created some bigger ones with $n=115, 135, 150$ nodes using the same procedure as above with the only 
difference that 25\%, 50\% and 75\% of the maximum number of edges were taken. 
\end{itemize}

To generate all these input graphs the {\tt R/igraph} package was used with the appropriate functions. 

\subsubsection{Real-world graphs}
The graphs have been used are benchmark graphs, the first set of the graphs taken from UCINET software datasets\footnote{https://sites.google.com/site/ucinetsoftware/datasets/}, the other graphs are well-known graphs from Network Repository\footnote{http://networkrepository.com/}.


\subsection{Discussion on results with random graphs}

\subsubsection{General observations}
Before the discussion of the results for the different types of random graphs, a general overview can be summarized as follows.
\begin{itemize}
\item The exact solutions were found less than half of the cases, and this is caused by running out of time. 

\item Both of the two versions of the greedy algorithm (Algorithm \ref{alg:3}) were able to finish their execution below 2 seconds for each and every tested random graph. Their execution times were roughly equal. The version which takes pair of vertices into account ($AddOne=0$) either gives better or equal upper bound compared with the version taking one vertex ($AddOne=1$). 

\item For the locally greedy algorithm (Algorithm \ref{alg:4}) the execution time was less than 0.2 seconds for each and every tested graph. Thus, this algorithm is the fastest one on average.

\item Not surprisingly, as the number of nodes increases the geodetic number and its computational time increases as well (for the same density). 
Taking the averages for the graphs with the same density we can conclude that the geodetic number is decreasing as the density is increasing, while the computational effort is maximal when the density is $0.4$, and minimal when it is $0.8$.

\end{itemize}

\subsubsection{Erd\H os-R\'enyi random graphs}
The results for the Erd\H os-R\'enyi graphs are reported in Table \ref{table:er}. The ILP solver was able to find the optimal solution within the time limit for 22 instances (45\% of the cases).

The greedy algorithm missed the optimal solution in 28 cases. So it reported the optimal value as upper bound in 43\% of the cases. For 20 graphs it missed the optimum with one more vertex, and for 7 graphs with two more vertices in the minimal geodetic set.
On the other hand, it found a better solution in one case than the reported upper bound by Gurobi. Comparing the two versions of the greedy: in 38 cases they reported the same values, {\tt AddOne} resulted in better upper bounds only for 2 input graphs, while the default version gave better upper bounds in 9 cases.

The locally greedy algorithm reported the optimal solution in 22 cases (45\% of the cases). In 16 cases the upper bound missed the optimal solution with one more vertex, in 6 cases with two more vertices and in 5 cases with three or more vertices.

Comparing the greedy algorithm with the locally greedy algorithm: the two algorithms reported the same upper bound in 32 cases. In 7 cases locally greedy algorithm gave better upper bound, while in 10 cases the greedy algorithm was closer to the optimal solution.

The last line of Table \ref{table:er} reports the mean values for the obtained bounds of the geodetic number as well as the average execution times. Although the differences are really small, the greedy algorithm is the best in getting good upper bounds, whereas the locally greedy method was the fastest one for the Erd\H os-R\'enyi random graphs.

\subsubsection{Watts-Strogatz}
The results for the Watts-Strogatz graphs are shown in Table \ref{table:ws}. 
The exact method ran out of time in 23 instances, so it was able to solve the problem in 53\% of the cases. 

The greedy algorithm obtained the same value as the exact method in 29\% of the cases (for 14 graphs). When it did not find the optimal (or best reported) value, it gave +1 for 25 graphs and +2 for 10 graphs.

 In 38 cases the two versions of the greedy algorithm reported the same values, {\tt AddOne} resulted in better upper bounds in 1 case only, while the default version gave better upper bounds in 10 times.
 
  The locally greedy algorithm reported the optimal solution in 8 cases. In 22 cases the upper bound missed the optimal solution with one more vertex, in 12 cases with two more vertices and in 7 cases with three or more vertices. 
  
 The greedy algorithm and the locally algorithm reported the same upper bound in 29 cases, the locally gave better upper bound in 3 cases, while the default version gave better upper bound in 17 cases.
 
 Regarding the average performance, which is reported in the last line of Table \ref{table:ws}, the default greedy algorithm was the best with respect to the upper bound, while the locally greedy approach was the quickest for the Watts-Strogatz random graphs.
 
\subsubsection{Barab\'asi-Albert}
Finally, for the Barab\'asi-Albert graphs the computational results are reported in Table \ref{table:ba}.
Gurobi was able to find the optimal solution within the time limit for 25 instances (51\% of the cases).
The greedy algorithm was not able to obtain the same value as the exact method in 31 cases, meaning a 37\% success rate.
For 21 graphs it reported +1, for 8 graphs +2 and for one graph +3 compared to the value obtained by the exact method.
For one graph instance it found better upper bound than Gurobi. 
By comparing the two versions of the greedy algorithm: in 29 cases they reported the same values, {\tt AddOne} resulted in better upper bound in 2 cases, while the default version gave better upper bounds in 18 times.

The locally greedy algorithm achieved the optimal solution in 10 cases. In 19 cases the upper bound missed the optimal solution with one more vertex, in 10 cases with two more vertices and in 10 cases with three or more vertices.

The default version of greedy algorithms and locally greedy algorithm reported the same upper bound in 25 case. In 4 cases locally gave better upper bound, whereas in 20 cases the default version was better.

The last line of Table \ref{table:ba} reports the average performances. Similarly to the other two types of random graphs, the default greedy algorithm is the best one for obtaining upper bound for the geodetic number, and the locally greedy algorithm can do this in the fastest time for the Barab\'asi-Albert random graphs.

\begin{table}[H]
\centering
\caption{Numerical results for the Erd\H os-R\'enyi random graphs, time is given in seconds}
\begin{tabular}{cccccccccc}
\hline
\multicolumn{2}{c}{graph} & \multicolumn{2}{c}{exact} & \multicolumn{2}{c}{greedy} & \multicolumn{2}{c}{greedy ({\tt AddOne})} & \multicolumn{2}{c}{locally greedy}\\
$n$ & $m$  & value& time& value& time& value& time& value & time \\
\hline
10 & 9 & 4 & 0.004 & 4 & 0.004 & 4 & 0.003 & \cellcolor[HTML]{C0C0C0}4 & \cellcolor[HTML]{C0C0C0}0.001 \\
10 & 18 & 4 & 0.004 & 4 & 0.004 & 4 & 0.004 & 4 & 0.004 \\
10 & 30 & 4 & 0.007 & 4 & 0.002 & 4 & 0.002 & 4 & 0.002 \\
10 & 36 & 3 & 0.011 & 4 & 0.004 & 4 & 0.004 & \cellcolor[HTML]{C0C0C0}3&\cellcolor[HTML]{C0C0C0}0.004 \\
20 & 37 & 5 & 0.063 & 5 & 0.009 & \cellcolor[HTML]{C0C0C0}5 & 0.008& 6& 0.004 \\
20 & 76 & 4 & 0.073 & 4 & 0.011 & 4 & 0.011 & \cellcolor[HTML]{C0C0C0}4 & \cellcolor[HTML]{C0C0C0}0.006 \\
20 & 114 & 4 & 0.767 & 5 & 0.01 & 5 & 0.011 & \cellcolor[HTML]{C0C0C0}5 & \cellcolor[HTML]{C0C0C0}0.007 \\
20 & 152 & 3 & 0.197 & 3 & 0.011 & 3 & 0.012 & \cellcolor[HTML]{C0C0C0}3 & \cellcolor[HTML]{C0C0C0}0.006 \\
30 & 87 & 6 & 0.906 & 7 & 0.018 & 8 & 0.017 & \cellcolor[HTML]{C0C0C0}7 &\cellcolor[HTML]{C0C0C0} 0.011 \\
30 & 174 & 6 & 3.258 & 8 & 0.019 & 8 & 0.019 & \cellcolor[HTML]{C0C0C0}6 & \cellcolor[HTML]{C0C0C0}0.013 \\
30 & 261 & 4 & 2.198 & 5 & 0.016 & 5 & 0.015 & \cellcolor[HTML]{C0C0C0}5 & \cellcolor[HTML]{C0C0C0}0.011 \\
30 & 348 & 4 & 3.259 & 4 & 0.02 & 4 & 0.018 & \cellcolor[HTML]{C0C0C0}4 & \cellcolor[HTML]{C0C0C0}0.009 \\
40 & 156 & 7 & 76.422 & 7 & 0.03 & \cellcolor[HTML]{C0C0C0}7 & 0.027 & 9 & 0.011 \\
40 & 312 & 6 & 89.274 & 8 & 0.031 & 7 & 0.031 & \cellcolor[HTML]{C0C0C0}7 & \cellcolor[HTML]{C0C0C0}0.016 \\
40 & 468 & 4 & 17.424 & 6 & 0.039 & 6 & 0.037 & \cellcolor[HTML]{C0C0C0}6 & \cellcolor[HTML]{C0C0C0}0.011 \\
40 & 624 & 3 & 4.086 & 4 & 0.036 & 4 & 0.037 & \cellcolor[HTML]{C0C0C0}4 & \cellcolor[HTML]{C0C0C0}0.013 \\
50 & 245 & 7 & 208.332 & 9 & 0.056 & \cellcolor[HTML]{C0C0C0}9 & 0.051 & 10 & 0.023 \\
50 & 490 & 6 & 1169.8 & \cellcolor[HTML]{C0C0C0}7 & 0.059 & 8 & 0.052 & 8 & 0.018 \\
50 & 735 & 5 & 832.257 & 6 & 0.068 & 6 & 0.059 & \cellcolor[HTML]{C0C0C0}5 & \cellcolor[HTML]{C0C0C0}0.014 \\
50 & 1000 & 3 & 18.231 & 4 & 0.061 & 4 & 0.057 & \cellcolor[HTML]{C0C0C0}4 & \cellcolor[HTML]{C0C0C0}0.016 \\
60 & 354 & $\leq$   9 & $>$ 3600 & \cellcolor[HTML]{C0C0C0}10 & 0.092 & 11 & 0.083 & 11 & 0.028 \\
60 & 708 & $\leq$   7 & $>$ 3600 & 8 & 0.106 & 8 & 0.096 & \cellcolor[HTML]{C0C0C0}8 & \cellcolor[HTML]{C0C0C0}0.025 \\
60 & 1062 & $\leq$   5 & $>$ 3600 & 6 & 0.111 & 6 & 0.097 & \cellcolor[HTML]{C0C0C0}5 & \cellcolor[HTML]{C0C0C0}0.024 \\
60 & 1416 & 4 & 429.412 & 4 & 0.104 & 4 & 0.101 & \cellcolor[HTML]{C0C0C0}4 & \cellcolor[HTML]{C0C0C0}0.016 \\
70 & 483 & $\leq$   8 & $>$ 3600 & 10 & 0.131 & \cellcolor[HTML]{C0C0C0}10 & 0.121 & 12 & 0.033 \\
70 & 966 & $\leq$   7 & $>$ 3600 & 9 & 0.164 & 9 & 0.139 & \cellcolor[HTML]{C0C0C0}9 & \cellcolor[HTML]{C0C0C0}0.038 \\
70 & 1449 & $\leq$   5 & $>$ 3600 & 6 & 0.156 & 7 & 0.152 & \cellcolor[HTML]{C0C0C0}6 & \cellcolor[HTML]{C0C0C0}0.031 \\
70 & 1932 & 4 & 663.129 & 4 & 0.147 & 4 & 0.142 & \cellcolor[HTML]{C0C0C0}4 & \cellcolor[HTML]{C0C0C0}0.025 \\
80 & 632 & $\leq$   9 & $>$ 3600 & \cellcolor[HTML]{C0C0C0}9 & 0.193 & 10 & 0.181 & 13 & 0.048 \\
80 & 1264 & $\leq$   8 & $>$ 3600 & 9 & 0.235 & 10 & 0.208 & \cellcolor[HTML]{C0C0C0}9 & \cellcolor[HTML]{C0C0C0}0.047 \\
80 & 1896 & $\leq$   6 & $>$ 3600 & 6 & 0.242 & 6 & 0.215 & \cellcolor[HTML]{C0C0C0}6 & \cellcolor[HTML]{C0C0C0}0.039 \\
80 & 2528 & $\leq$   4 & $>$ 3600 & 4 & 0.225 & 4 & 0.214 & \cellcolor[HTML]{C0C0C0}4 & \cellcolor[HTML]{C0C0C0}0.029 \\
90 & 801 & $\leq$   10 & $>$ 3600 & \cellcolor[HTML]{C0C0C0} 10 & 0.277 &13 & 0.261 & 15 & 0.638 \\
90 & 1602 & $\leq$   8 & $>$ 3600 & 9 & 0.331 & 10 & 0.295 & \cellcolor[HTML]{C0C0C0}9 & \cellcolor[HTML]{C0C0C0}0.052 \\
90 & 2403 & $\leq$   5 & $>$ 3600 & 6 & 0.347 & 6 & 0.316 & \cellcolor[HTML]{C0C0C0}5 & \cellcolor[HTML]{C0C0C0}0.041 \\
90 & 3204 & $\leq$   4 & $>$ 3600 & 5 & 0.319 & 5 & 0.302 & \cellcolor[HTML]{C0C0C0}5 & \cellcolor[HTML]{C0C0C0}0.041 \\
100 & 990 & $\leq$   12 & $>$ 3600 & \cellcolor[HTML]{C0C0C0}11 & 0.375 & 14 & 0.348 &15 &0.076 \\
100 & 1980 & $\leq$   9 & $>$ 3600 & 9 & 0.449 & 9 & 0.387 & \cellcolor[HTML]{C0C0C0}9 & \cellcolor[HTML]{C0C0C0}0.061 \\
100 & 2970 & $\leq$   6 & $>$ 3600 & 6 & 0.479 & 6 & 0.423 & \cellcolor[HTML]{C0C0C0}6 & \cellcolor[HTML]{C0C0C0}0.047 \\
100 & 3960 & $\leq$   4 & $>$ 3600 & 4 & 0.433 & 4 & 0.425 & \cellcolor[HTML]{C0C0C0}4 & \cellcolor[HTML]{C0C0C0}0.038 \\
115 & 1638 & $\leq$   13 & $>$ 3600 & 15 & 0.627 & 15 & 0.572 & \cellcolor[HTML]{C0C0C0}14 & \cellcolor[HTML]{C0C0C0}0.101 \\
115 & 3277 & $\leq$   7 & $>$ 3600 & 8 & 0.746 & 8 & 0.648 & \cellcolor[HTML]{C0C0C0}8 & \cellcolor[HTML]{C0C0C0}0.074 \\
115 & 4916 & $\leq$   5 & $>$ 3600 & 5 & 0.704 & 5 & 0.672 & \cellcolor[HTML]{C0C0C0}5 & \cellcolor[HTML]{C0C0C0}0.068 \\
135 & 2261 & $\leq$   14 & $>$ 3600 & 15 & 0.943 & \cellcolor[HTML]{C0C0C0}14 & 0.823 & 16 & 0.152 \\
135 & 4522 & $\leq$   8 & $>$ 3600 & 8 & 1.107 & 8 & 0.953 & \cellcolor[HTML]{C0C0C0}8 & \cellcolor[HTML]{C0C0C0}0.107 \\
135 & 6783 & $\leq$   4 & $>$ 3600 & 5 & 1.101 & 5 & 1.042 & \cellcolor[HTML]{C0C0C0}5 & \cellcolor[HTML]{C0C0C0}0.088 \\
150 & 2793 & $\leq$   15 & $>$ 3600 & 16 & 1.393 & 16 & 1.154 & \cellcolor[HTML]{C0C0C0}16 & \cellcolor[HTML]{C0C0C0}0.182 \\
150 & 5587 & $\leq$   8 & $>$ 3600 & 8 & 1.528 & 8 & 1.356 & \cellcolor[HTML]{C0C0C0}8 & \cellcolor[HTML]{C0C0C0}0.134 \\
150 & 8381 & $\leq$   5 & $>$ 3600 & 5 & 1.563 & 5 & 1.465 & \cellcolor[HTML]{C0C0C0}5 & \cellcolor[HTML]{C0C0C0}0.107 \\
\hline
\multicolumn{2}{c}{avarge} & 6.224 & 2055.492 & \cellcolor[HTML]{C0C0C0}6.898 & 0.309 & 7.122 & 0.279 & 7.184 & 0.053\\
\hline
\end{tabular}
\label{table:er}
\end{table}

\begin{table}[H]
\centering
\caption{Numerical results for the Watts-Strogatz random graphs, time is given in seconds}
\begin{tabular}{cccccccccc}
\hline
\multicolumn{2}{c}{graph} & \multicolumn{2}{c}{exact} & \multicolumn{2}{c}{greedy} & \multicolumn{2}{c}{greedy ({\tt AddOne})} & \multicolumn{2}{c}{locally greedy}\\
$n$&$m$&value& time& value & time & value & time & value &time \\\hline
10 & 9 & 5 & 0.008 & 5 & 0.003 & 5 & 0.003 & \cellcolor[HTML]{C0C0C0}5 & \cellcolor[HTML]{C0C0C0}0.002 \\
10 & 18 & 3 & 0.013 & 3 & 0.004 & 3 & 0.004 & \cellcolor[HTML]{C0C0C0}3 &\cellcolor[HTML]{C0C0C0} 0.003 \\
10 & 30 & 3 & 0.012 & 3 & 0.004 & 3 & 0.004 & \cellcolor[HTML]{C0C0C0}3 & \cellcolor[HTML]{C0C0C0}0.003 \\
10 & 36 & 2 & 0.007 & 2 & 0.001 & 2 & 0.001 & 2 & 0.001 \\
20 & 37 & 5 & 0.102 & 6 & 0.011 & 6 & 0.008 & \cellcolor[HTML]{C0C0C0}6 & \cellcolor[HTML]{C0C0C0}0.007 \\
20 & 76 & 5 & 0.512 & 6 & 0.011 & 6 & 0.011 & \cellcolor[HTML]{C0C0C0}6 & \cellcolor[HTML]{C0C0C0}0.008 \\
20 & 114 & 4 & 0.477 & 4 & 0.009 & 4 & 0.008 & \cellcolor[HTML]{C0C0C0}4 & \cellcolor[HTML]{C0C0C0}0.006 \\
20 & 152 & 4 & 0.728 & \cellcolor[HTML]{C0C0C0}4 & 0.011 & 5 & 0.011 & 5 & 0.007 \\
30 & 87 & 7 & 2.294 & 7 & 0.019 & 7\cellcolor[HTML]{C0C0C0} & 0.018 & 9 & 0.013 \\
30 & 174 & 6 & 6.423 & \cellcolor[HTML]{C0C0C0}6 & 0.02 & 6 & 0.02 & 7 & 0.011 \\
30 & 261 & 5 & 3.852 & 5 & 0.022 & 5 & 0.019 & \cellcolor[HTML]{C0C0C0}5 &\cellcolor[HTML]{C0C0C0} 0.012 \\
30 & 348 & 4 & 1.052 & 4 & 0.019 & 4 & 0.018 & \cellcolor[HTML]{C0C0C0}4 & \cellcolor[HTML]{C0C0C0}0.006 \\
40 & 156 & 7 & 20.866 & 9 & 0.036 & \cellcolor[HTML]{C0C0C0}9 & 0.032 & 10 & 0.018 \\
40 & 312 & 6 & 106.294 & \cellcolor[HTML]{C0C0C0}6 & 0.035 & 9 & 0.032 & 8 & 0.019 \\
40 & 468 & 4 & 35.792 & 4 & 0.038 & \cellcolor[HTML]{C0C0C0}4 & 0.035 & 5 & 0.015 \\
40 & 624 & 3 & 2.963 & 4 & 0.035 & 5 & 0.034 & \cellcolor[HTML]{C0C0C0}3 & \cellcolor[HTML]{C0C0C0}0.011 \\
50 & 245 & 7 & 82.019 & 8 & 0.054 &\cellcolor[HTML]{C0C0C0} 8 & 0.047 & 11 & 0.023 \\
50 & 490 & $\leq$ 7 & $>$ 3600 & 9 & 0.067 & 9 & 0.058 & \cellcolor[HTML]{C0C0C0}9 & \cellcolor[HTML]{C0C0C0}0.024 \\
50 & 735 & 5 & 255.426 & 6 & 0.072 & 6 & 0.063 & \cellcolor[HTML]{C0C0C0}6 & \cellcolor[HTML]{C0C0C0}0.019 \\
50 & 1000 & 4 & 4.742 & 5 & 0.059 & 5 & 0.058 & \cellcolor[HTML]{C0C0C0}5 & \cellcolor[HTML]{C0C0C0}0.017 \\
60 & 354 & 9 & 771.2 & \cellcolor[HTML]{C0C0C0}10 & 0.094 & 11 & 0.085 & 12 & 0.031 \\
60 & 708 & $\leq$ 7 & $>$ 3600 & 9 & 0.102 & 9 & 0.094 & \cellcolor[HTML]{C0C0C0}9 & \cellcolor[HTML]{C0C0C0}0.031 \\
60 & 1062 & 5 & 561.631 & 7 & 0.107 & 8 & 0.1 & \cellcolor[HTML]{C0C0C0}6 & \cellcolor[HTML]{C0C0C0}0.026 \\
60 & 1416 & 4 & 10.923 & 5 & 0.097 & 5 & 0.095 & \cellcolor[HTML]{C0C0C0}5 & \cellcolor[HTML]{C0C0C0}0.021 \\
70 & 483 & $\leq$ 8 & $>$ 3600 & 10 & 0.138 & \cellcolor[HTML]{C0C0C0}9 & 0.126 & 11 & 0.035 \\
70 & 966 & $\leq$ 7 & $>$ 3600 & 9 & 0.161 & 9 & 0.137 & \cellcolor[HTML]{C0C0C0}9 & \cellcolor[HTML]{C0C0C0}0.037 \\
70 & 1449 & $\leq$ 5 & $>$ 3600 & 6 & 0.157 & 6 & 0.141 & \cellcolor[HTML]{C0C0C0}6 &\cellcolor[HTML]{C0C0C0} 0.026 \\
70 & 1932 & 4 & 28.026 & 5 & 0.148 & 5 & 0.137 & \cellcolor[HTML]{C0C0C0}5 & \cellcolor[HTML]{C0C0C0}0.024 \\
80 & 632 & $\leq$ 9 & $>$ 3600 & 10 & 0.208 & \cellcolor[HTML]{C0C0C0}10 & 0.183 & 13 & 0.047 \\
80 & 1264 & $\leq$ 8 & $>$ 3600 & \cellcolor[HTML]{C0C0C0}9 & 0.238 & 10 & 0.207 & 10 & 0.048 \\
80 & 1896 & $\leq$ 5 & $>$ 3600 & 7 & 0.239 & 7 & 0.216 & \cellcolor[HTML]{C0C0C0}7 & \cellcolor[HTML]{C0C0C0}0.042 \\
80 & 2528 & 4 & 390.748 & 5 & 0.213 & 5 & 0.208 & \cellcolor[HTML]{C0C0C0}5 & \cellcolor[HTML]{C0C0C0}0.028 \\
90 & 801 & $\leq$ 10 & $>$ 3600 & \cellcolor[HTML]{C0C0C0}11 & 0.276 & 13 & 0.265 & 13 & 0.058 \\
90 & 1602 & $\leq$ 8 & $>$ 3600 & \cellcolor[HTML]{C0C0C0}9 & 0.336 & 9 & 0.286 & 10 & 0.061 \\
90 & 2403 & $\leq$ 6 & $>$ 3600 & 7 & 0.344 & 7 & 0.311 & \cellcolor[HTML]{C0C0C0}7 & \cellcolor[HTML]{C0C0C0}0.049 \\
90 & 3204 & 4 & 106.156 & 6 & 0.327 & 6 & 0.302 & \cellcolor[HTML]{C0C0C0}5 & \cellcolor[HTML]{C0C0C0}0.041 \\
100 & 990 & $\leq$ 11 & $>$ 3600 & \cellcolor[HTML]{C0C0C0}11 & 0.371 & 13 & 0.343 & 15 & 0.082 \\
100 & 1980 & $\leq$ 8 & $>$ 3600 & 9 & 0.472 & 9 & 0.387 & \cellcolor[HTML]{C0C0C0}9 & \cellcolor[HTML]{C0C0C0}0.068 \\
100 & 2970 & $\leq$ 6 & $>$ 3600 & 7 & 0.452 & 7 & 0.417 & \cellcolor[HTML]{C0C0C0}7 & \cellcolor[HTML]{C0C0C0}0.059 \\
100 & 3960 & 4 & 350.92 & 6 & 0.415 & 6 & 0.398 & \cellcolor[HTML]{C0C0C0}6 & \cellcolor[HTML]{C0C0C0}0.043 \\
115 & 1638 & $\leq$   13 & $>$ 3600 & 14 & 0.614 & \cellcolor[HTML]{C0C0C0}14 & 0.574 & 15 & 0.106 \\
115 & 3277 & $\leq$ 8 & $>$ 3600 & 9 & 0.715 & 9 & 0.648 & \cellcolor[HTML]{C0C0C0}9 & \cellcolor[HTML]{C0C0C0}0.091 \\
115 & 4916 & $\leq$ 5 & $>$ 3600 & 6 & 0.693 & 6 & 0.645 & \cellcolor[HTML]{C0C0C0}6 & \cellcolor[HTML]{C0C0C0}0.071 \\
135 & 2261 & $\leq$ 14 & $>$ 3600 & 15 & 0.927 & 16 & 0.832 & \cellcolor[HTML]{C0C0C0}15 & \cellcolor[HTML]{C0C0C0}0.135 \\
135 & 4522 & $\leq$ 7 & $>$ 3600 & 9 & 1.112 & 9 & 0.952 & \cellcolor[HTML]{C0C0C0}9 & \cellcolor[HTML]{C0C0C0}0.116 \\
135 & 6783 & $\leq$ 5 & $>$ 3600 & 6 & 1.065 & 6 & 0.997 & \cellcolor[HTML]{C0C0C0}6 & \cellcolor[HTML]{C0C0C0}0.086 \\
150 & 2793 & $\leq$ 13 & $>$ 3600 & \cellcolor[HTML]{C0C0C0}14 & 1.287 & 16 & 1.164 & 15 & 0.183 \\
150 & 5587 & $\leq$ 8 & $>$ 3600 & 8 & 1.497 & \cellcolor[HTML]{C0C0C0}8 & 1.338 & 9 & 0.141 \\
150 & 8381 & $\leq$ 5 & $>$ 3600 & 6 & 1.502 & 6 & 1.411 & \cellcolor[HTML]{C0C0C0}6 & \cellcolor[HTML]{C0C0C0}0.106 \\
\hline
\multicolumn{2}{c}{average} & 6.265 & 1745.78 & \cellcolor[HTML]{C0C0C0}7.163 & 0.303 & 7.45 & 0.28 & 7.673 & 0.043\\
\hline
\end{tabular}
\label{table:ws}
\end{table}

\begin{table}[H]
\centering
\caption{Numerical results for the Barab\'asi-Albert random graphs, time is given in seconds}
\begin{tabular}{cccccccccc}
\hline
\multicolumn{2}{c}{graph} & \multicolumn{2}{c}{exact} & \multicolumn{2}{c}{greedy} & \multicolumn{2}{c}{greedy ({\tt AddOne})} & \multicolumn{2}{c}{locally greedy}\\
$n$&$m$&value& time& value & time & value & time & value &time \\\hline
10 & 9 & 6 & 0.004 & 8 & 0.004 & 6 & 0.004 & \cellcolor[HTML]{C0C0C0}6 & \cellcolor[HTML]{C0C0C0}0.002 \\
10 & 18 & 4 & 0.007 & 4 & 0.004 & 4 & 0.004 & \cellcolor[HTML]{C0C0C0}4 & \cellcolor[HTML]{C0C0C0}0.001 \\
10 & 30 & 4 & 0.007 & 4 & 0.004 & 4 & 0.004 & \cellcolor[HTML]{C0C0C0}4 & \cellcolor[HTML]{C0C0C0}0.002 \\
10 & 36 & 3 & 0.009 & 3 & 0.002 & 3 & 0.001 & \cellcolor[HTML]{C0C0C0}3 & \cellcolor[HTML]{C0C0C0}0.001 \\
20 & 37 & 7 & 0.042 & 7 & 0.012 & \cellcolor[HTML]{C0C0C0}7 & 0.011 & 8 & 0.008 \\
20 & 76 & 5 & 0.058 & 6 & 0.011 & 7 & 0.009 & \cellcolor[HTML]{C0C0C0}6 & \cellcolor[HTML]{C0C0C0}0.007 \\
20 & 114 & 4 & 0.176 & 5 & 0.012 & 5 & 0.012 & \cellcolor[HTML]{C0C0C0}5 & \cellcolor[HTML]{C0C0C0}0.007 \\
20 & 152 & 3 & 0.046 & 3 & 0.008 & 3 & 0.007 & \cellcolor[HTML]{C0C0C0}3 & \cellcolor[HTML]{C0C0C0}0.005 \\
30 & 87 & 9 & 0.269 & \cellcolor[HTML]{C0C0C0}10 & 0.021 & 12 & 0.02 & 11 & 0.011 \\
30 & 174 & 6 & 0.382 & 7 & 0.02 & 8 & 0.017 & \cellcolor[HTML]{C0C0C0}7 & \cellcolor[HTML]{C0C0C0}0.012 \\
30 & 261 & 5 & 0.995 & 5 & 0.016 & 5 & 0.015 & \cellcolor[HTML]{C0C0C0}5 & \cellcolor[HTML]{C0C0C0}0.007 \\
30 & 348 & 3 & 0.724 & 4 & 0.019 & 4 & 0.019 & \cellcolor[HTML]{C0C0C0}4 & \cellcolor[HTML]{C0C0C0}0.011 \\
40 & 156 & 10 & 20.117 & \cellcolor[HTML]{C0C0C0}11 & 0.032 & 12 & 0.033 & 12 & 0.016 \\
40 & 312 & 7 & 4.014 & 10 & 0.038 & 10 & 0.035 & \cellcolor[HTML]{C0C0C0}10 & \cellcolor[HTML]{C0C0C0}0.02 \\
40 & 468 & 5 & 9.844 & 7 & 0.037 & 7 & 0.034 & \cellcolor[HTML]{C0C0C0}7 & \cellcolor[HTML]{C0C0C0}0.017 \\
40 & 624 & 4 & 1.955 & 4 & 0.036 & 4 & 0.035 & \cellcolor[HTML]{C0C0C0}4 & \cellcolor[HTML]{C0C0C0}0.013 \\
50 & 245 & 11 & 6.754 & 12 & 0.064 & \cellcolor[HTML]{C0C0C0}12 & 0.051 & 13 & 0.026 \\
50 & 490 & 8 & 197.846 & 10 & 0.063 & \cellcolor[HTML]{C0C0C0}9 & 0.056 & 10 & 0.022 \\
50 & 735 & 5 & 58.891 & 7 & 0.067 & 8 & 0.061 & \cellcolor[HTML]{C0C0C0}6 & \cellcolor[HTML]{C0C0C0}0.019 \\
50 & 1000 & 4 & 12.052 & 6 & 0.059 & 6 & 0.053 & \cellcolor[HTML]{C0C0C0}5 & \cellcolor[HTML]{C0C0C0}0.014 \\
60 & 354 & $\leq$ 11 & $>$ 3600 & 12 & 0.097 & \cellcolor[HTML]{C0C0C0}12 & 0.092 & 16 & 0.037 \\
60 & 708 & $\leq$ 8 & $>$ 3600 & 10 & 0.103 & 10 & 0.096 & \cellcolor[HTML]{C0C0C0}9 & \cellcolor[HTML]{C0C0C0}0.031 \\
60 & 1062 & 6 & 645.853 & \cellcolor[HTML]{C0C0C0}6 & 0.101 & 7 & 0.094 & 7 & 0.028 \\
60 & 1416 & 4 & 26.435 & 4 & 0.092 & 4 & 0.089 & \cellcolor[HTML]{C0C0C0}4 & \cellcolor[HTML]{C0C0C0}0.018 \\
70 & 483 & $\leq$ 11 & $>$ 3600 & \cellcolor[HTML]{C0C0C0}11 & 0.142 & 14 & 0.129 & 18 & 0.048 \\
70 & 966 & $\leq$ 9 & $>$ 3600 & 11 & 0.164 & 11 & 0.139 & \cellcolor[HTML]{C0C0C0}11 & \cellcolor[HTML]{C0C0C0}0.043 \\
70 & 1449 & $\leq$ 6 & $>$ 3600 & 6 & 0.151 & 6 & 0.135 & \cellcolor[HTML]{C0C0C0}6 & \cellcolor[HTML]{C0C0C0}0.025 \\
70 & 1932 & 4 & 67.771 & 5 & 0.146 & 5 & 0.137 & \cellcolor[HTML]{C0C0C0}5 & \cellcolor[HTML]{C0C0C0}0.027 \\
80 & 632 & $\leq$ 11 & $>$ 3600 & \cellcolor[HTML]{C0C0C0}11 & 0.207 & 12 & 0.187 & 18 & 0.068 \\
80 & 1264 & $\leq$ 9 & $>$ 3600 & 9 & 0.224 & \cellcolor[HTML]{C0C0C0}9 & 0.203 & 11 & 0.051 \\
80 & 1896 & $\leq$ 6 & $>$ 3600 & 7 & 0.231 & 7 & 0.208 & \cellcolor[HTML]{C0C0C0}7 & \cellcolor[HTML]{C0C0C0}0.041 \\
80 & 2528 & 4 & 132.305 & 5 & 0.217 & 5 & 0.207 & \cellcolor[HTML]{C0C0C0}5 & \cellcolor[HTML]{C0C0C0}0.032 \\
90 & 801 & $\leq$ 13 & $>$ 3600 & \cellcolor[HTML]{C0C0C0}14 & 0.305 & 17 & 0.283 & 20 & 0.081 \\
90 & 1602 & $\leq$ 9 & $>$ 3600 & \cellcolor[HTML]{C0C0C0}10 & 0.315 & 11 & 0.281 & 11 & 0.064 \\
90 & 2403 & $\leq$ 6 & $>$ 3600 & 7 & 0.334 & 8 & 0.311 & \cellcolor[HTML]{C0C0C0}7 & \cellcolor[HTML]{C0C0C0}0.047 \\
90 & 3204 & 4 & 713.099 & 5 & 0.305 & 5 & 0.283 & \cellcolor[HTML]{C0C0C0}5 & \cellcolor[HTML]{C0C0C0}0.042 \\
100 & 990 & $\leq$ 15 & $>$ 3600 & \cellcolor[HTML]{C0C0C0}14 & 0.392 & 18 & 0.363 & 18 & 0.101 \\
100 & 1980 & $\leq$ 9 & $>$ 3600 & \cellcolor[HTML]{C0C0C0}10 & 0.421 & 13 & 0.378 & 13 & 0.081 \\
100 & 2970 & $\leq$ 7 & $>$ 3600 & 8 & 0.435 & 8 & 0.402 & \cellcolor[HTML]{C0C0C0}8 & \cellcolor[HTML]{C0C0C0}0.063 \\
100 & 3960 & $\leq$ 4 & $>$ 3600 & 5 & 0.416 & 5 & 0.387 & \cellcolor[HTML]{C0C0C0}5 & \cellcolor[HTML]{C0C0C0}0.052 \\
115 & 1638 & $\leq$ 15 & $>$ 3600 & \cellcolor[HTML]{C0C0C0}16 & 0.626 & 18 & 0.566 & 18 & 0.121 \\
115 & 3277 & $\leq$ 8 & $>$ 3600 & 8 & 0.673 & \cellcolor[HTML]{C0C0C0}8 & 0.608 & 10 & 0.081 \\
115 & 4916 & $\leq$ 5 & $>$ 3600 & 5 & 0.648 & \cellcolor[HTML]{C0C0C0}5 & 0.632 & 6 & 0.062 \\

135 & 2261 & $\leq$ 15 & $>$ 3600 & \cellcolor[HTML]{C0C0C0}15 & 0.938 & 18 & 0.828 & 21 & 0.176 \\
135 & 4522 & $\leq$ 9 & $>$ 3600 & 10 & 1.064 & 11 & 0.927 & \cellcolor[HTML]{C0C0C0}10 & \cellcolor[HTML]{C0C0C0}0.115 \\
135 & 6783 & $\leq$ 5 & $>$ 3600 & 6 & 1.053 & \cellcolor[HTML]{C0C0C0}6 & 0.963 & 7 & 0.106 \\
150 & 2793 & $\leq$ 15 & $>$ 3600 & 17 & 1.282 & \cellcolor[HTML]{C0C0C0}17 & 1.192 & 20 & 0.221 \\
150 & 5587 & $\leq$ 9 & $>$ 3600 & \cellcolor[HTML]{C0C0C0}9 & 1.443 & 11 & 1.291 & 10 & 0.137 \\
150 & 8381 & $\leq$ 6 & $>$ 3600 & 6 & 1.415 & 7 & 1.342 & \cellcolor[HTML]{C0C0C0}6 & \cellcolor[HTML]{C0C0C0}0.118 \\
\hline
\multicolumn{2}{c}{average} & 7.27 &1802.034 & \cellcolor[HTML]{C0C0C0}8.061 & 0.3 & 8.653 & 0.27 & 9.081 & 0.048\\
\hline
\end{tabular}
\label{table:ba}
\end{table}

\subsection{Discussion of the results on larger graph instances}

It is obvious from the values reported in Table \ref{table:real}  that ILP can take many hours to get the exact geodetic number for graphs with thousands of nodes and edges, while our proposed algorithms were able to obtain acceptable upper bounds in reasonable time. Note that for this set of graphs we did not use any time limit for Gurobi. Even for the largest graph instance (ia-email-univ) both variants of the greedy algorithm were able to report a slightly worse upper bound than the exact value of in less than 700 seconds. Although the locally greedy algorithm was the fastest, for the larger graphs it missed the upper bound by a much larger margin than the greedy method. This trend can also be seen in the last line of Table \ref{table:real}, where the average perofmrances are reported.

\begin{table}
\centering
\caption{Numerical results for real-world graphs, time is given in seconds, unless indicated otherwise}
\begin{tabular}{ccccccccccc}
\hline
\multicolumn{3}{c}{graph} &
  \multicolumn{2}{c}{exact} &
  \multicolumn{2}{c}{greedy} &
  \multicolumn{2}{c}{greedy (\tt{AddOne})} &
  \multicolumn{2}{c}{locally greedy} \\ 
name &
  \multicolumn{1}{c}{$n$} &
  \multicolumn{1}{c}{$m$} &
  \multicolumn{1}{c}{value} &
  \multicolumn{1}{c}{time} &
  \multicolumn{1}{c}{value} &
  \multicolumn{1}{c}{time} &
  \multicolumn{1}{c}{value} &
  \multicolumn{1}{c}{time} &
  \multicolumn{1}{c}{value} &
  \multicolumn{1}{c}{time} \\\hline
karate           & 34   & 78   & 16  & 0.06 & 16  & 0.024 & 16  & 0.019 & 16  & 0.015  \\
mexican          & 35   & 117  & 7   & 0.21 & 7   & 0.025 & 8   & 0.023 & 9   & 0.011  \\
sawmill          & 36   & 62   & 14  & 0.09 & 14  & 0.022 & 14  & 0.019 & 15  & 0.015  \\
chesapeake       & 39   & 170  & 5   & 0.12 & 5   & 0.026 & 5   & 0.023 & 5   & 0.008  \\ \hline
ca-netscience    & 379  & 914  & 253 & 37 m  & 256 & 21.6  & 260 & 20.8  & 264 & 14.1   \\
bio-celegans     & 453  & 2025 & 172 & 1 h   & 183 & 40.8  & 188 & 34.3  & 225 & 14.8   \\
rt-twitter-copen & 761  & 1029 & 459 & 6 h   & 459 & 101.7 & 459 & 103.4 & 490 & 112.6  \\
soc-wiki-vote    & 889  & 2914 & 275 & 14 h  & 276 & 236.4 & 277 & 232.0 & 409 & 120.5  \\
ia-email-univ    & 1133 & 5451 & 244 & 16 h  & 248 & 698.6 & 250 & 677.9 & 464 & 269.0  \\\hline
average   & & &   160.56&  4 h & 162.67 & 122.14 & 164.11  & 118.72 & 210.78 & 58.99 \\\hline
\end{tabular}
\label{table:real}
\end{table}

\section{Conclusions}
Given the fact that the graph geodetic number problem is NP-hard, it is desirable to establish efficient algorithmic upper bounds 
which can provide feasible solutions of acceptable quality and reasonable running time.
We have proposed greedy type approaches and experimentally shown that these algorithms can determine such upper bounds and their running time can be small fraction of the time required to obtain the exact geodetic number.

\section*{Acknowledgments} 
The research was supported by the European Union, co-financed by the European Social Fund (EFOP-3.6.3-VEKOP-16-2017-00002)
and by the Ministry for Innovation and Technology, Hungary (grant TUDFO/47138-1/2019-ITM).

\bibliographystyle{spmpsci}


\end{document}